\documentclass[11pt,
showpacs,
preprintnumbers,amsmath,amssymb,
aps,prd,nofootinbib,eqsecnum,a4paper]{revtex4}
\usepackage{epsfig}
\usepackage{graphicx,epsf}
\usepackage{color}
\usepackage{bm}
\usepackage{psfrag}
\usepackage[symbol]{footmisc}
\usepackage{tensor}
\usepackage{hyperref}
\def\beq{\begin{equation}}
\def\eeq{\end{equation}}
\def\bea{\begin{eqnarray}}
\def\eea{\end{eqnarray}}
\def\bi{\begin{itemize}}
\def\ei{\end{itemize}}
\def\cs2{c_{\rm{s}}^2}
\def \beg {\begin{enumerate}}
\def \en {\end{enumerate}}

\def\M0{{\cal M}_0}

\begin{document}

\title{A Natural Inflation inspired model}

\author{Gabriel Germ\'an\footnote[1]{\href{mailto:gabriel@icf.unam.mx}{gabriel@icf.unam.mx}}}
\affiliation{
$^2$Instituto de Ciencias F\'{i}sicas, Universidad Nacional Aut\'{o}noma de M\'{e}xico,\\Av. Universidad S/N. Cuernavaca, Morelos, 62251, M\'{e}xico}
%

\begin{abstract}
We propose a modification of the Natural Inflation (NI) potential in such a way that the spontaneous symmetry breaking scale $f$ can take values less than one (in Planck units). The proposed potential seems simple enough, however, its consequences are difficult to calculate analytically. Therefore, we illustrate the feasibility of the model by considering some numerical examples that easily satisfy the conditions imposed on the observables $n_s$ and $r$ by the most recent observations, while at the same time maintaining the number of e-folds during the inflationary epoch within the expected range.
  
\end{abstract}


\maketitle

\section {\bf Introduction}\label{INT}

Natural Inflation (NI) \cite{Freese:1990rb} is an inflationary model proposed in 1990 which over the years has been of interest to the community working on the subject due to its simplicity and its motivation by attractive physical concepts such as  the protection of the flatness of the potential by a shift symmetry, spontaneous symmetry breaking and mass generation for the resulting Goldstone Boson through non-perturbative effects. However, the most recent observations, mainly due to the Planck satellite, have ended up discarding the model despite numerous efforts to keep it viable. In the present work we propose a slight modification of the model that nevertheless is capable of satisfying all the requirements for an acceptable inflation while keeping the model close to its original spirit. For this we consider the two equivalent versions of NI (with the inflaton rolling away from the origin in one case and towards the origin in the other) and generalize them by means of a power over the cosine term. After doing this, it is evident that the part of the potential relevant to inflation can be obtained by a simple translation by $f\pi/2$  which places, e.g., the flat part of the potential at the origin. At the origin, both the first and second derivatives of the potential are zero and it is shown that there are power values such that the resulting potential is capable of satisfying all inflationary requirements while producing an acceptable number of e-folds during inflation. 

The organization of the article is as follows: In Section \ref{NIR} we briefly discuss the NI model with emphasis on certain aspects not covered in the literature and with the intention of eventually reproducing them in the modified model. In particular, we show how the potential of NI transitions to a power law potential when the decay constant $f$ approaches infinity. In this way we easily understand Fig.~8 of Planck's article  \cite{Akrami:2018odb} where we can see how the solution $r(n_s, N_ {ke})$ for NI ends in the monomial  $\phi^2$ i.e., the observed effect in  \cite{Akrami:2018odb} is easily explained now because for large $f$ the potential of NI tends to the monomial reaching it in the limit $f\rightarrow\infty$. This limit is equivalent to $r\rightarrow 4\delta_{n_s}$ where $\delta_{n_s}$ is defined as $\delta_{n_s}\equiv 1-n_s$. For $r>4\delta_{n_s}$ $f$ loses its character as a symmetry breaking scale, the $\cos(\frac{\phi}{f})$ term becomes a $\cosh(\frac{\phi}{f})$ and the NI potential transitions to a new potential which, although theoretically interesting, is not acceptable phenomenologically.

In Section \ref{NII} we briefly present simple arguments for obtaining the model that modifies NI. Here, we treat separately the two equivalent ways in which the NI potential is usually discussed. We see that the modification of both versions lead to the same potential (given by Eq.~\eqref{pot}) which is the one we study in the numerical calculations of Section \ref{VIA} where we show that the model is capable of satisfying the inflationary requirements particularly for a sub-Planckian field $vev$ with the scale $f$ taking values as low as $10^{-8}$ in Planck units. Low energy inflation models have been discussed previously beginning with \cite{Knox:1992iy} as well as modifications of natural inflation involving higher powers of the inflaton (see e.g., \cite{Kinney:1995cc}, \cite{Kinney:1995ki}). However, we believe that the  modification strategy presented here for NI is particularly simple and straightforward and can be applied to other models of inflation \cite{German:2021abc} with promising results. Finally, we briefly present our conclusions in Section \ref{CON}.
\section {\bf Natural Inflation revisited}\label{NIR} 

In Fig.~8 of the Planck 2018 collaboration paper {\it Constraints on inflation} \cite{Akrami:2018odb}  (reproduced here as Fig.\,\ref{Plancky}) we can see that the (purple) lines for the tensor-to-scalar ratio $r$ as a function of the spectral index $n_s$  end in the monomial $\phi^2$. This is done for $N_{ke}=50,60$ e-folds of inflation, $N_{ke}$ denotes the number of e-folds from horizon crossing at $\phi_k$ to the end of inflation at $\phi_e$. In analogy with a similar study \cite{German:2021tqs} for the $\alpha$-attractors \cite{Kallosh:2013yoa} we show how this comes about by working out the Natural Inflation (NI) potential in terms of the observables $n_s$ and $r$ and from there we show how it smoothly transitions to the power law potential $V(\phi)=\frac{1}{2}m^2\phi^2$. 

\begin{figure}[tb]
\begin{center}
\includegraphics[width=10cm]{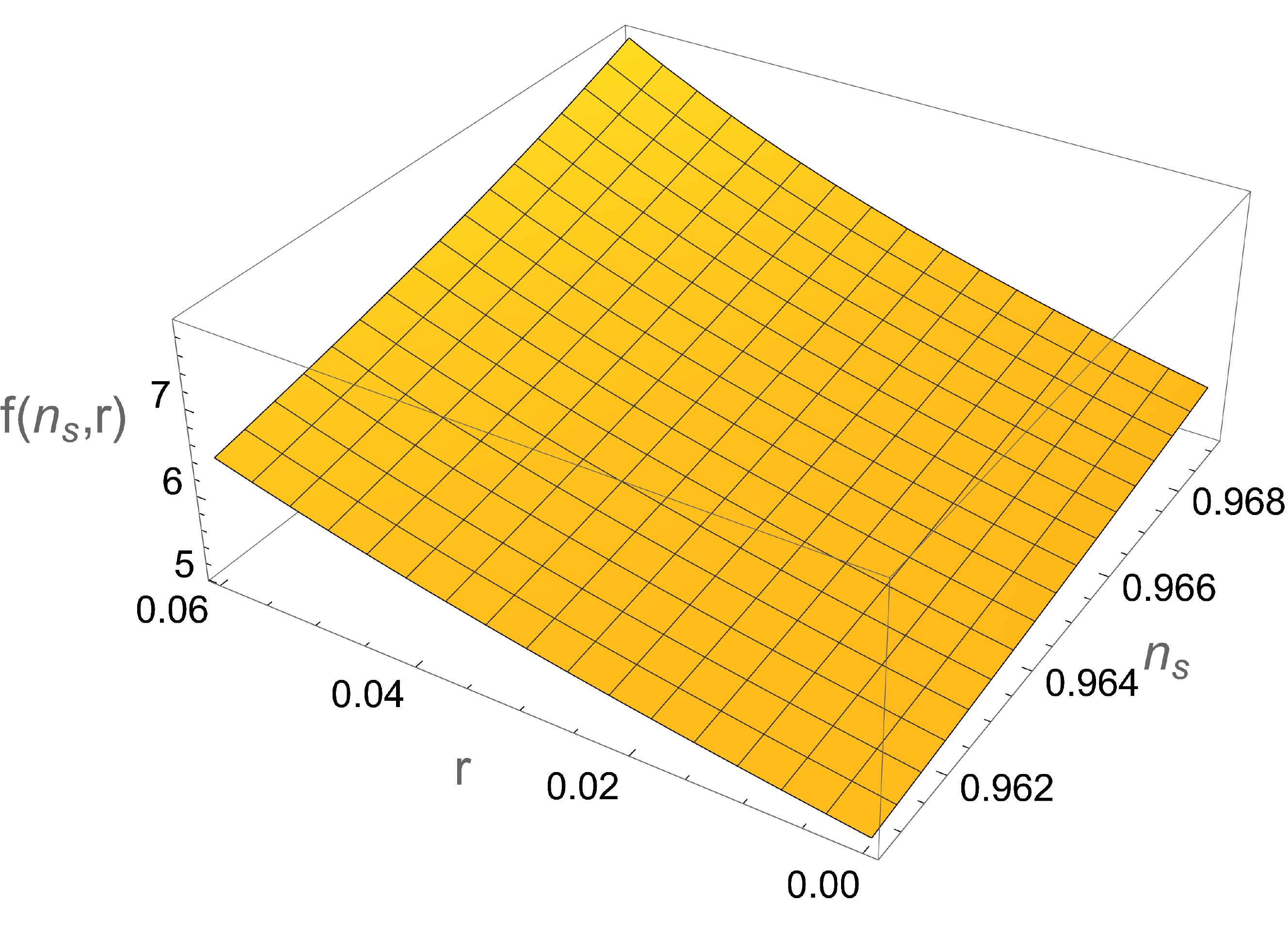}
\caption{\small Plot of the symmetry breaking scale $f$ given by Eq.~\eqref{f} as a function of the observables $n_s$ and $r$. For the range reported by the Planck 2018 Collaboration article \cite{Akrami:2018odb}, $n_s=0.9649\pm 0.0042$ and $r<0.06$, it is found that $f$ is bounded as $5 < f <8.1$.
}
\label{fnsr}
\end{center}
\end{figure}

The NI potential can be written in the form
\begin{equation}
V = V_0\left(1- \cos\left(\frac{\phi}{f}\right)\right), 
\label{potNIminus}
\end{equation}
where $f$ is the scale at which (global) symmetry breaking occurs giving rise to a  Goldstone boson while non-perturbative physics at temperatures lower than $f$ provides a potential and with it a small mass for the (originally massless) boson which becomes a pseudo-Nambu-Goldstone boson; the inflaton.
From the equation defining the tensor-to-scalar ratio $r=16\epsilon$ we solve for $\phi_k$, the value of the inflaton at horizon crossing
\begin{equation}
\cos\left(\frac{\phi_k}{f}\right) =1-\frac{16}{8+f^2\,r}.
\label{fik}
\end{equation}
We now substitute $\phi_k$ in the expression for the spectral index $n_{s} =1+2\eta -6\epsilon$, written in the form $\delta_{n_s}+2\eta-\frac{3}{8}r=0$, where $\delta_{n_s}$ is defined as $\delta_{n_s}\equiv 1-n_s$. The solution of this equation for the scale $f$  is
\begin{equation}
f=\frac{2}{\sqrt{4\delta_{n_s}-r}} ,  
\label{f}
\end{equation}
from where we learn that the maximum value $r$ can take is $4\delta_{n_s}$. We plot in Fig.~\ref{fnsr} the parameter $f$ as a function of $n_s$ and $r$ for the range of values reported by Planck  \cite{Akrami:2018odb} : $n_s=0.9649\pm 0.0042$ and $r<0.06$. Within this range we find that $f$ is constrained to  $5<f<8.1$.
Similarly, through the equation
\begin{figure}[tb]
\begin{center}
\includegraphics[trim = 0mm  0mm 1mm 1mm, clip, width=7.5cm, height=6cm]{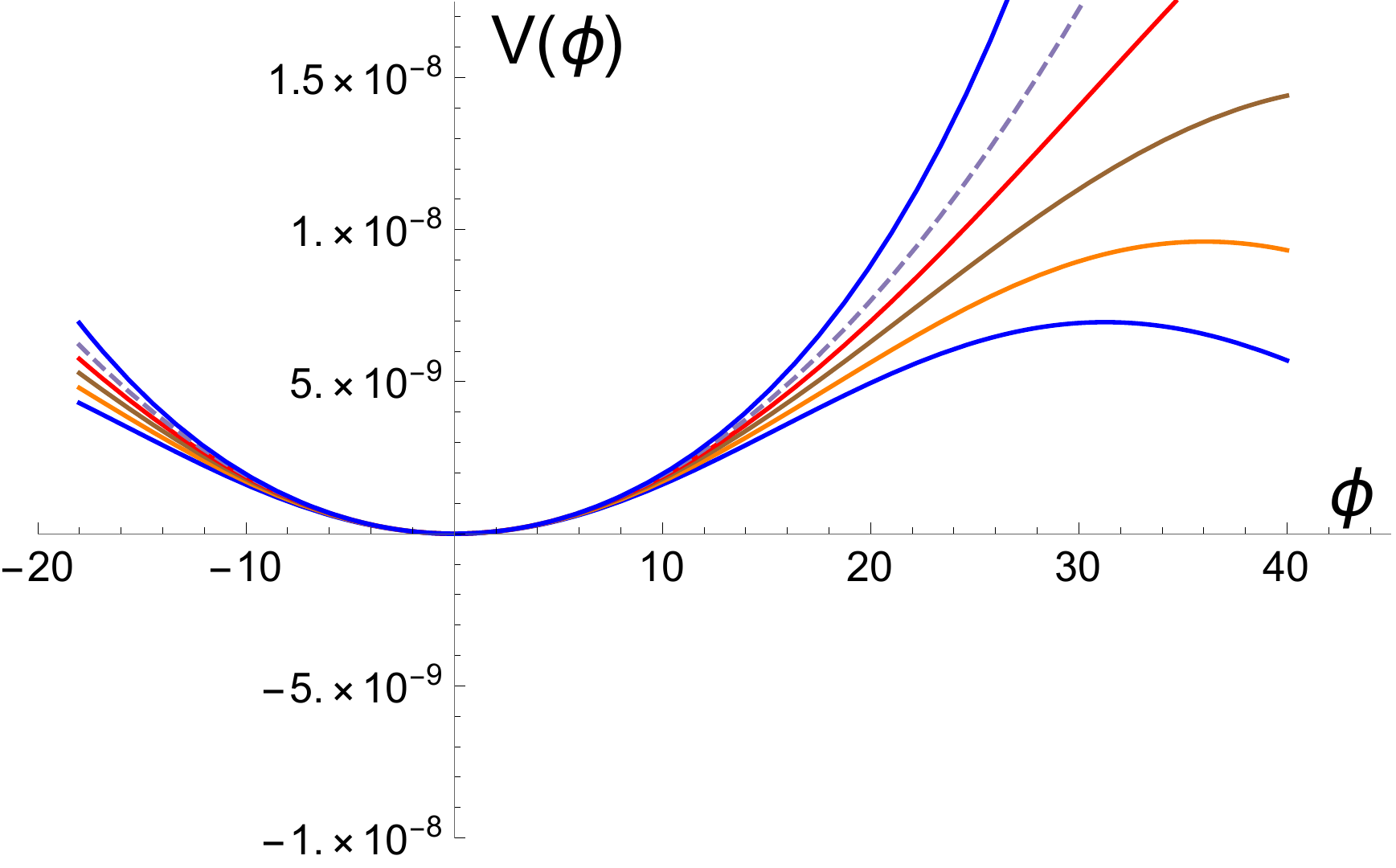}
\includegraphics[trim = 0mm  0mm 1mm 1mm, clip, width=7.5cm, height=6.cm]{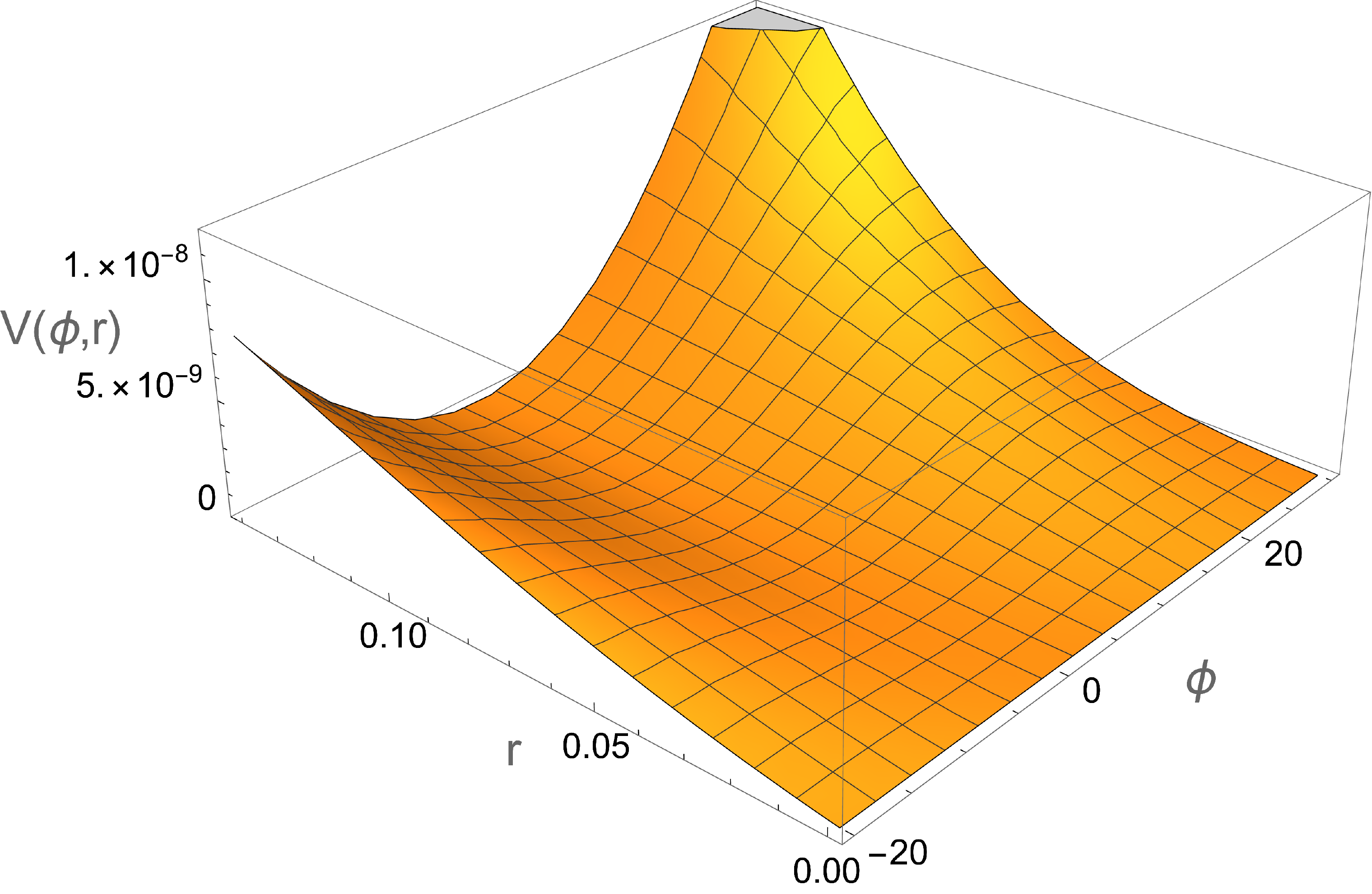}
\caption{\small  The l.h.s figure shows the potential \eqref{potnsr} for various values of $r$ reaching the maximum value $r=4\delta_{n_s}$ (dashed line) where the NI potential transitions to the power law potential $V(\phi)=m^2\phi^2/2$. The r.h.s. figure is a trivial extension in the $r$-direction. This transition, from the NI potential to a  power law potential, explains why, in the $n_s$-$r$ plane the lines $r(n_s,N_{ke})$ contain the monomial $\phi^2$ (see Figs.~\ref{cosas} and \ref{Plancky}). There is a line further back than the dashed line, this is because the NI potential for an imaginary argument transitions, passing the monomial, to a potential of the form given by Eq.~\eqref{potsinh}, however, this region is not acceptable phenomenologically. Fig.~\ref{cosas} shows that the monomial $\phi^2$ is the border between the potentials \eqref{potnsr} and \eqref{potsinh} valid for $r<4\delta_{n_s}$ and $r>4\delta_{n_s}$, respectively. }
\label{potNIPl}
\end{center}
\end{figure}
\begin{figure}[tb]
\begin{center}
\includegraphics[trim = 0mm  0mm 1mm 1mm, clip, width=7.5cm, height=6cm]{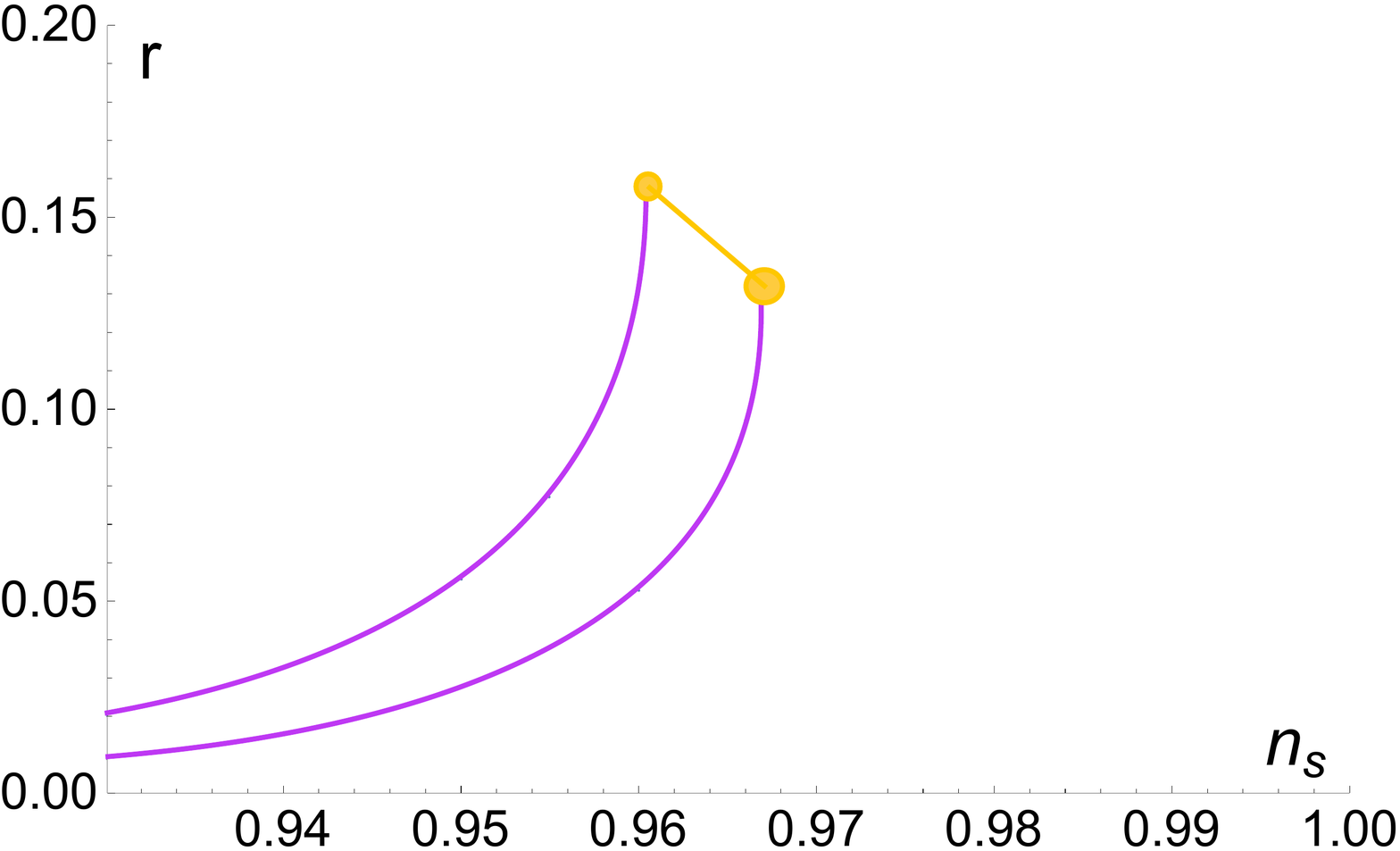}
\includegraphics[trim = 0mm  0mm 1mm 1mm, clip, width=7.5cm, height=6.cm]{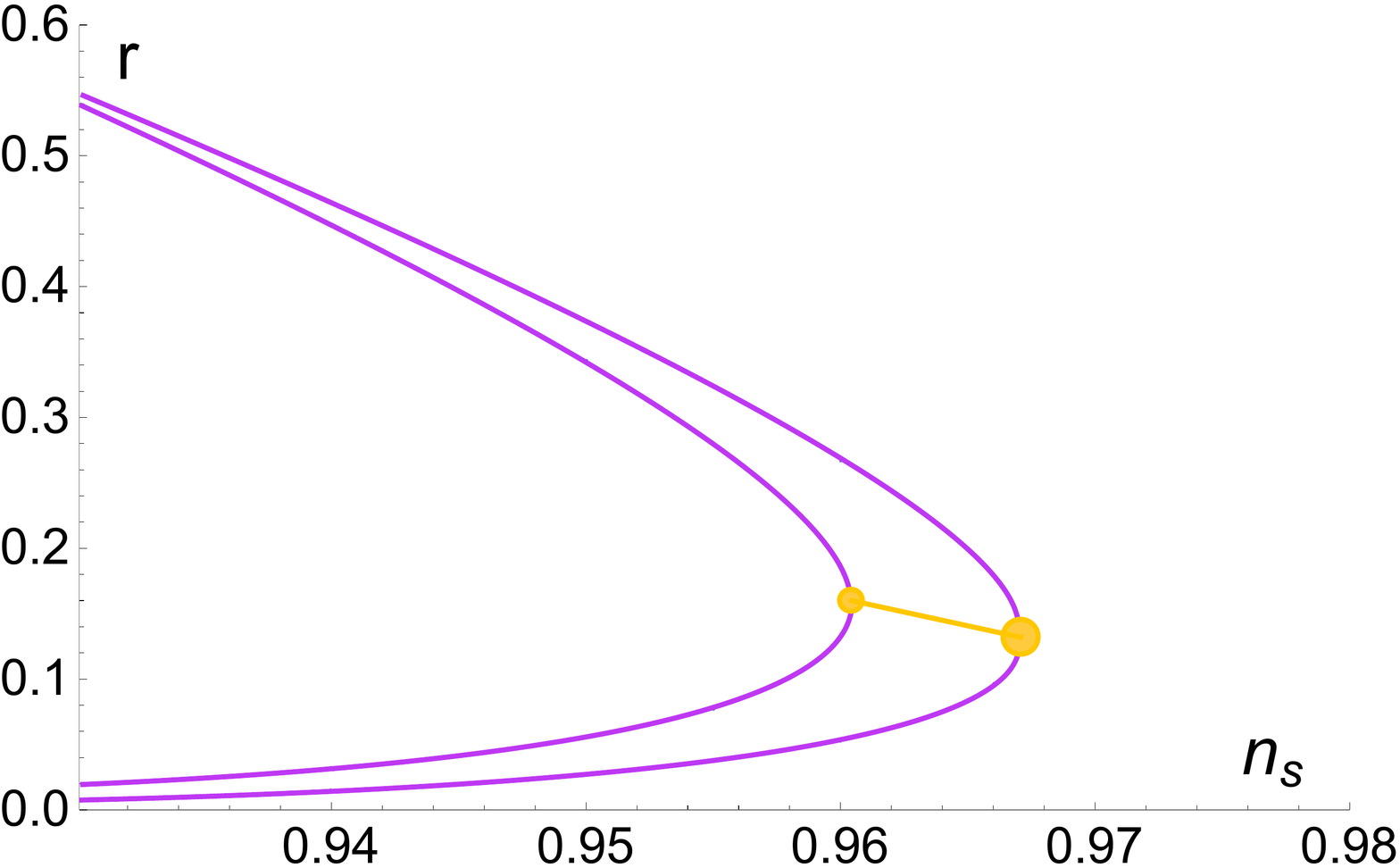}
\caption{\small In the l.h.s. figure we plot the tensor-to-scalar ratio $r$ as a function of $n_s$ for $N_{ke}=50$ (line ending in the small dot) and 60 e-folds of inflation. This figure is obtained by solving numerically Eq.~\eqref{Nkensr} and reproduces  the corresponding figure in Fig.~8 of Planck 2018 Collaboration article \cite{Akrami:2018odb}, duplicated here as Fig.~\ref{Plancky} below. In the r.h.s. (after the dots) the NI potential transitions to the potential given by Eq.~\eqref{potsinh} which is not of the NI-type with the constant $f$ losing its role as a symmetry breaking scale. Although the potential \eqref{potsinh} is a genuine inflationary potential it is not useful because phenomenologically is excluded. The monomial $\phi^2$ potential is thus the border between potentials \eqref{potnsr} and \eqref{potsinh}.  Looking at things the other way around we could understand the NI potential as originating from a $V=\bar V_0\left(-1+\cosh(\frac{\phi}{\bar f})\right)$ potential for $r>4\delta_{n_s}$ such that in the transition to $r<4\delta_{n_s}$, $\bar V_0\rightarrow -V_0$, $\bar f \rightarrow i f$ and $\cosh(\frac{\phi}{\bar f}) \rightarrow \cos(\frac{\phi}{f})$, generating the NI potential.
}
\label{cosas}
\end{center}
\end{figure}
\begin{figure}[tb]
\begin{center}
\includegraphics[width=12cm]{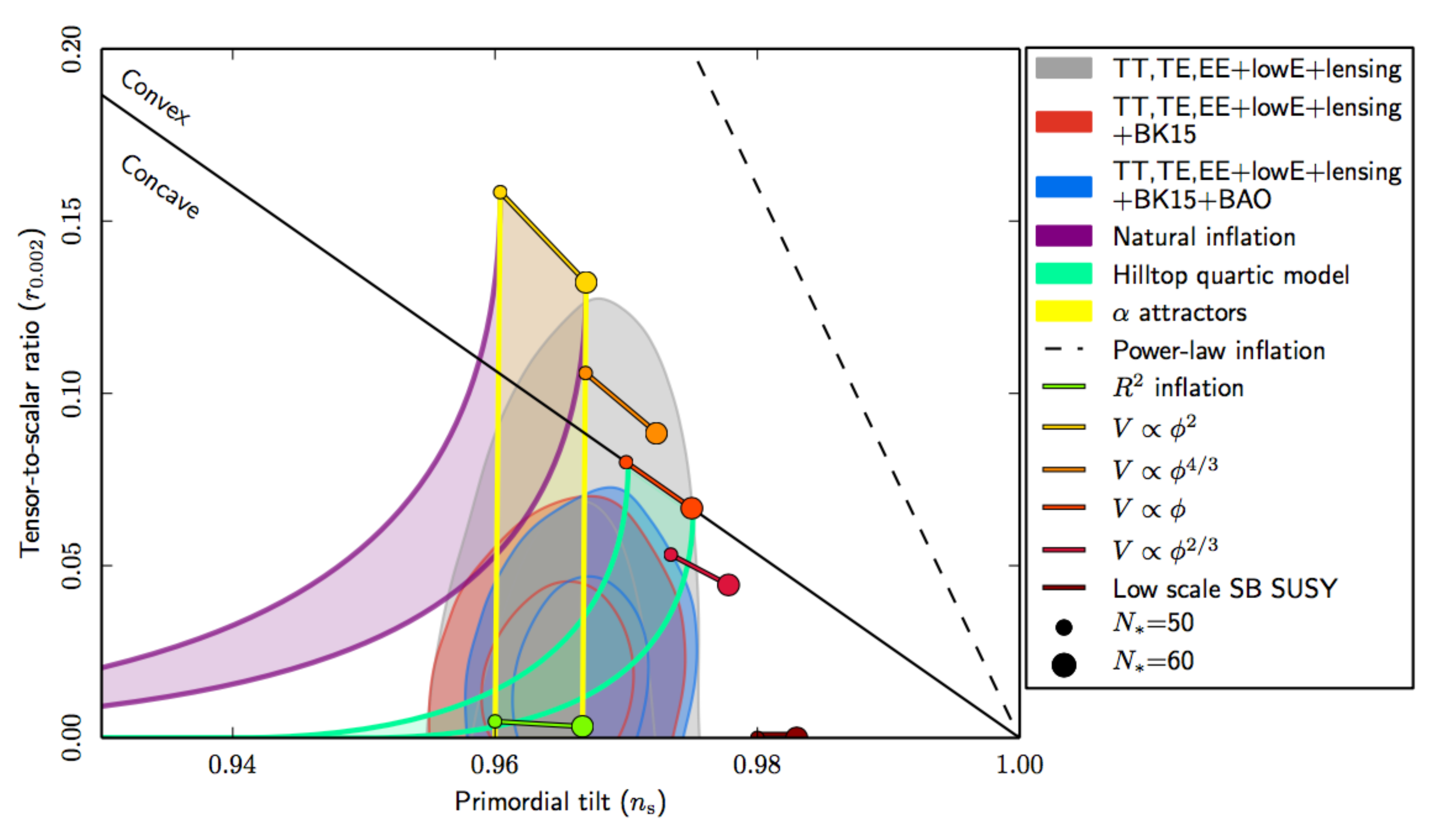}
\caption{\small We reproduce Fig.~8 of the Planck 2018 article \cite{Akrami:2018odb}  where monomial potentials are considered together with several other interesting models of inflation (see description in the right hand side panel of the figure). Natural Inflation ends in the $\phi^2$ monomial because the NI potential smoothly transitions to the power law potential $V=m^2\phi^2/2$ in the limit $f\rightarrow \infty$ as shown after Eq.~\eqref{potnsr} and illustrated in Fig.~\ref{potNIPl}.
}
\label{Plancky}
\end{center}
\end{figure}
\begin{equation}
A_s(k) =\frac{1}{24\pi ^{2}} \frac{V}{M^4\epsilon},
\label{IA} 
\end{equation}
we get
\begin{equation}
V_0=\frac{3A_s\pi^2r(8\delta_{n_s}-r)}{8(4\delta_{n_s}-r)} .
\label{V0}
\end{equation}
It follows that  the potential \eqref{potNIminus} can be written in terms of the observables $n_s$ and $r$ as 
\begin{equation}
V(\phi)=\frac{3A_s\pi^2r(8\delta_{n_s}-r)}{4(4\delta_{n_s}-r)}\sin^2\left(\frac{\sqrt{4\delta_{n_s}-r}}{4}\phi\right)\;, \quad\quad\quad r<4\delta_{n_s}.
\label{potnsr}
\end{equation}
In the limit when $r\rightarrow 4\delta_{n_s}$ or, more conveniently, $\delta_{n_s}\rightarrow r/4$  the potential \eqref{potnsr} transitions to
\begin{equation}
V(\phi)=\frac{3}{64}A_s\pi^2r^2\phi^2\;.
\label{mon}
\end{equation}
One can easily check that the potential given by Eq~\eqref{mon} corresponds exactly to the power law potential
\begin{equation}
V(\phi)=\frac{1}{2}m^2\phi^2\;,
\label{pl}
\end{equation}
after eliminating $m$ by means of Eq.~\eqref{IA}.
Thus NI smoothly transitions to a power law when $\delta_{n_s}\rightarrow r/4$ as can be seen in Fig.\,\ref{potNIPl} where we plot the potential   \eqref{potnsr} as a function of $\phi$ for various values of $r$ reaching $r=4\delta_{n_s}$ where the monomial $\phi^2$ occurs (dashed line). We can see, however, that a line further back than the dashed line has been drawn. This is because the NI potential for imaginary argument $(r>4\delta_{n_s})$ transitions to a potential of the form
 \begin{equation}
V(\phi)=\frac{3A_s\pi^2r(8\delta_{n_s}-r)}{4(r-4\delta_{n_s})}\sinh^2\left(\frac{\sqrt{r-4\delta_{n_s}}}{4}\phi\right)\;, \quad\quad\quad r>4\delta_{n_s}.
\label{potsinh}
\end{equation}
We can calculate an expression for the number of e-folds $N_{ke}$ in terms of $n_s$ and $r$ and from there attempt to solve for $r=r(n_s,N_{ke})$ however, the resulting equation for $N_{ke}$ (following from the original potential \eqref{potnsr}) is a trascendental equation
\begin{equation}
N_{ke}=\frac{4}{4\delta_{n_s}-r}\ln\left(\frac{8(8\delta_{n_s}-r)}{r(4(2+\delta_{n_s})-r)}\right)\;.
\label{Nkensr}
\end{equation}

Thus, we are satisfied with solving it numerically with the result shown in Fig.~\ref{cosas} reproducing the solution which appear in Fig.~8  of the Planck's article \cite{Akrami:2018odb} (see Fig.\,\ref{Plancky}) and extending it for the case $r>4\delta_{n_s}$ (in which case an overall minus sign should go into Eq.~\eqref{Nkensr} above).
This figure contains the monomial solutions of the power law potential as a transition between the potentials given by Eqs.~\eqref{potnsr} and \eqref{potsinh}. 

An alternative but equivalent way to show that the monomials are transition points of NI is by writing the number of e-folds $N_{ke}$ in terms of $f$ and $r$
\begin{equation}
N_{ke}=f^2\,\ln\left(\frac{2(8+f^2 r)}{r(1+2f^2)}\right)\;,
\label{Nkefr}
\end{equation}
and solving for $r$
\begin{equation}
r=\frac{16}{(1+2f^2)e^{\frac{N_{ke}}{f^2}}-2f^2}\;.
\label{rdeNkef}
\end{equation}
Taking the limit $f\rightarrow \infty$  (equivalently $\delta_{n_s}\rightarrow r/4$) we get
\begin{equation}
r=\frac{16}{2N_{ke}+1}\;.
\label{rlimit}
\end{equation}
This is precisely the relation (for the $p=2$ case) between $r$ and $N_{ke}$ for potentials of the power-law type $V_{pl}=\frac{1}{2}m^{4-p}\phi^p$ with $\phi_k=\frac{2\sqrt{2}\,p}{\sqrt{r}}$ and $\phi_e=\frac{p}{\sqrt{2}}$ from where it follows that
\begin{equation}
 r_{pl}=\frac{16p}{4N_{ke}+p}\,.
\label{rmon}
\end{equation}
These results are mainly of theoretical interest only because phenomenologically $r>4\delta_{n_s}$ violates bounds for $r$ and as we saw above, phenomenologically $f$ should be less than 8.1. 

For models of the $\alpha$-attractor type the potential is of the form $\tanh^p(\lambda \phi)$ and a situation similar to the one discussed above for NI occurs. For most values of the parameter $p$ the monomials are the end points of $\alpha$-attractors.  This is not the case for even-$p$ however, with the lines $r(n_s,N_{ke})$ continuing after they have encountered the monomial. This situation is also not favored by the data as in NI above and in both cases one can actually consider the monomials as the ending points of the {\it original model} for any $p$   \cite{Kallosh:2013yoa}, \cite{German:2021tqs}.
\section {\bf A Natural Inflation inspired model}\label{NII} 
\begin{figure}[t!]
\begin{center}
\includegraphics[trim = 0mm  0mm 1mm 1mm, clip, width=7.5cm, height=5.cm]{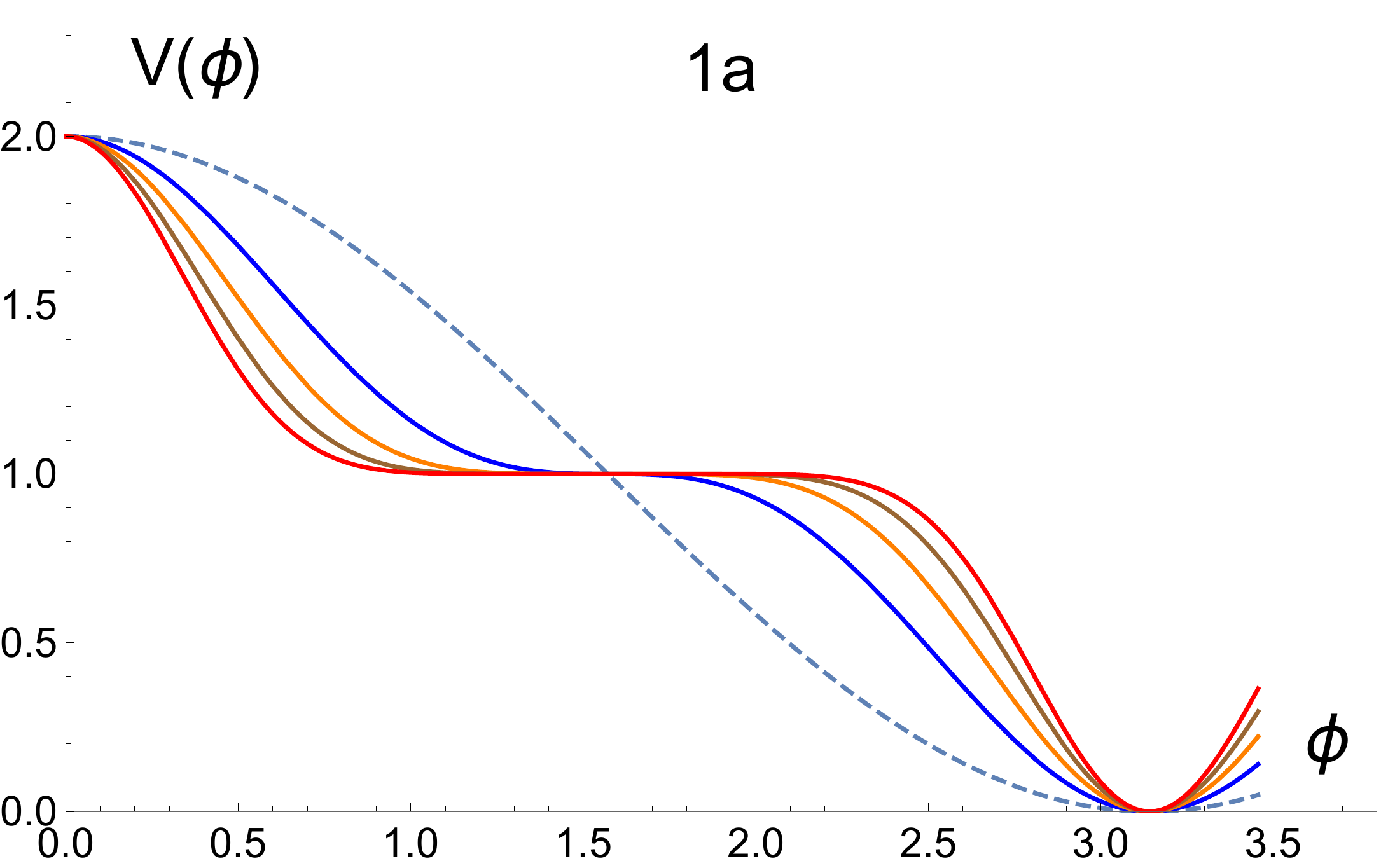}
\includegraphics[trim = 0mm  0mm 1mm 1mm, clip, width=7.5cm, height=5.cm]{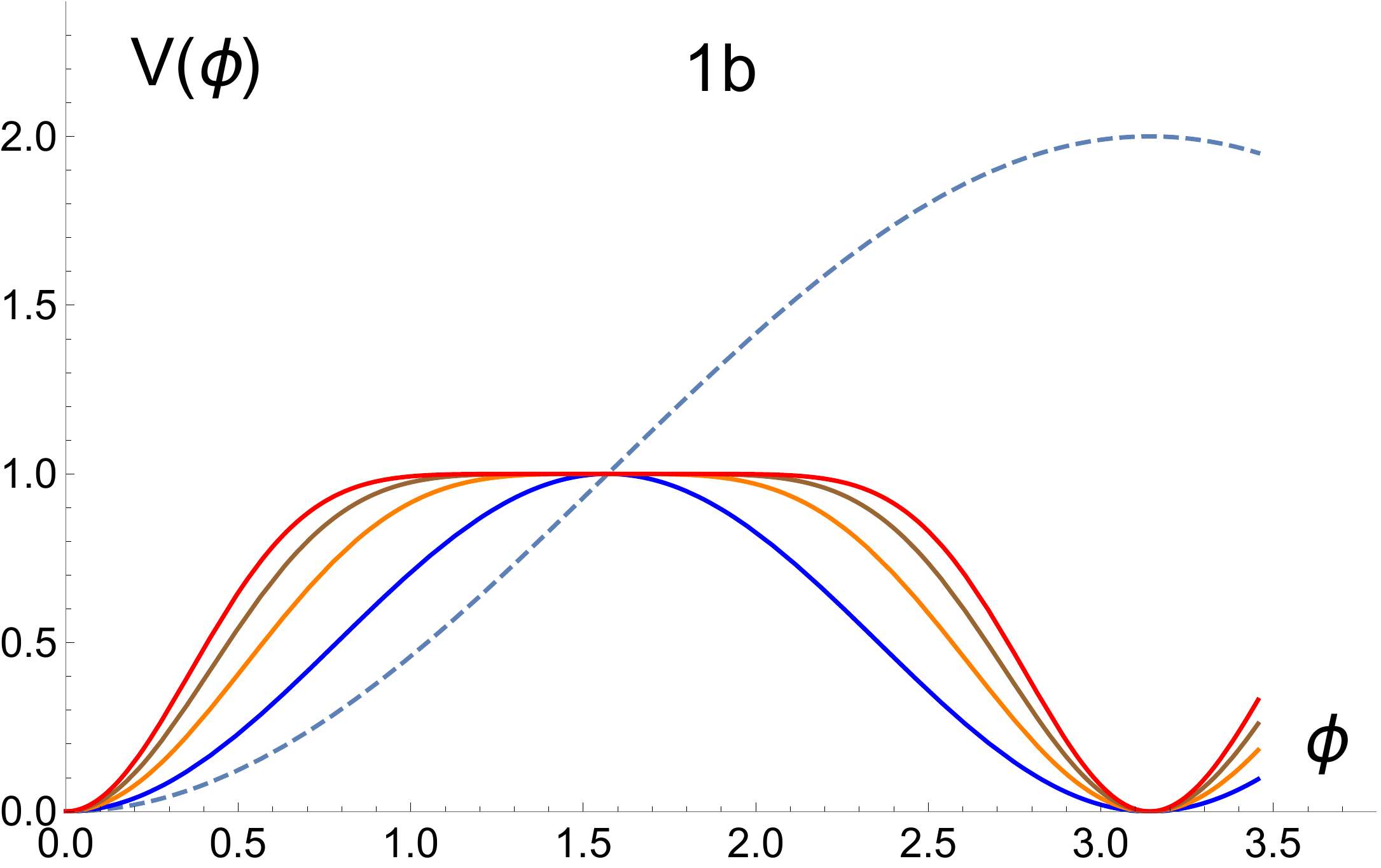}
\includegraphics[trim = 0mm  0mm 1mm 1mm, clip, width=7.5cm, height=5.cm]{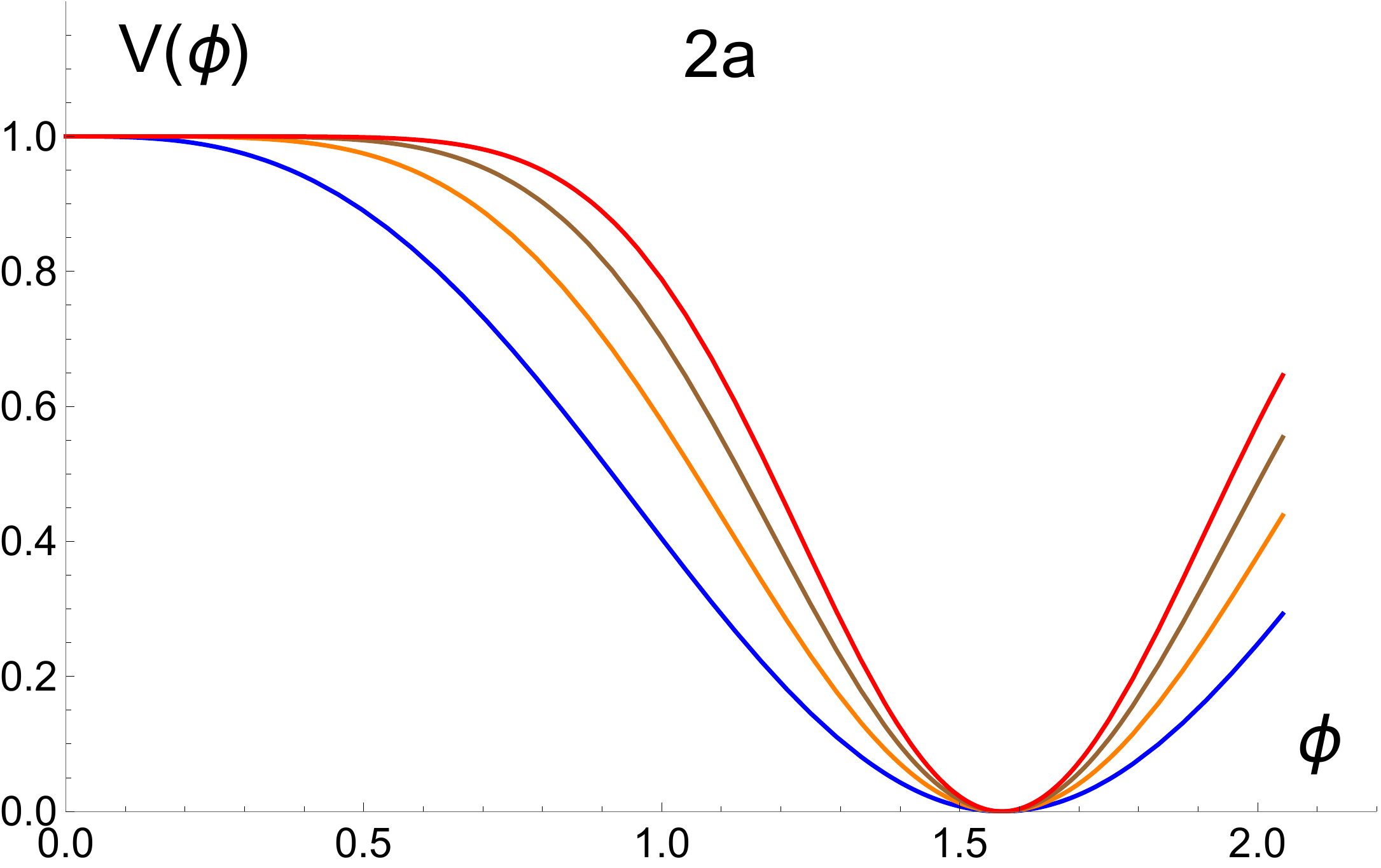}
\includegraphics[trim = 0mm  0mm 1mm 1mm, clip, width=7.5cm, height=5.cm]{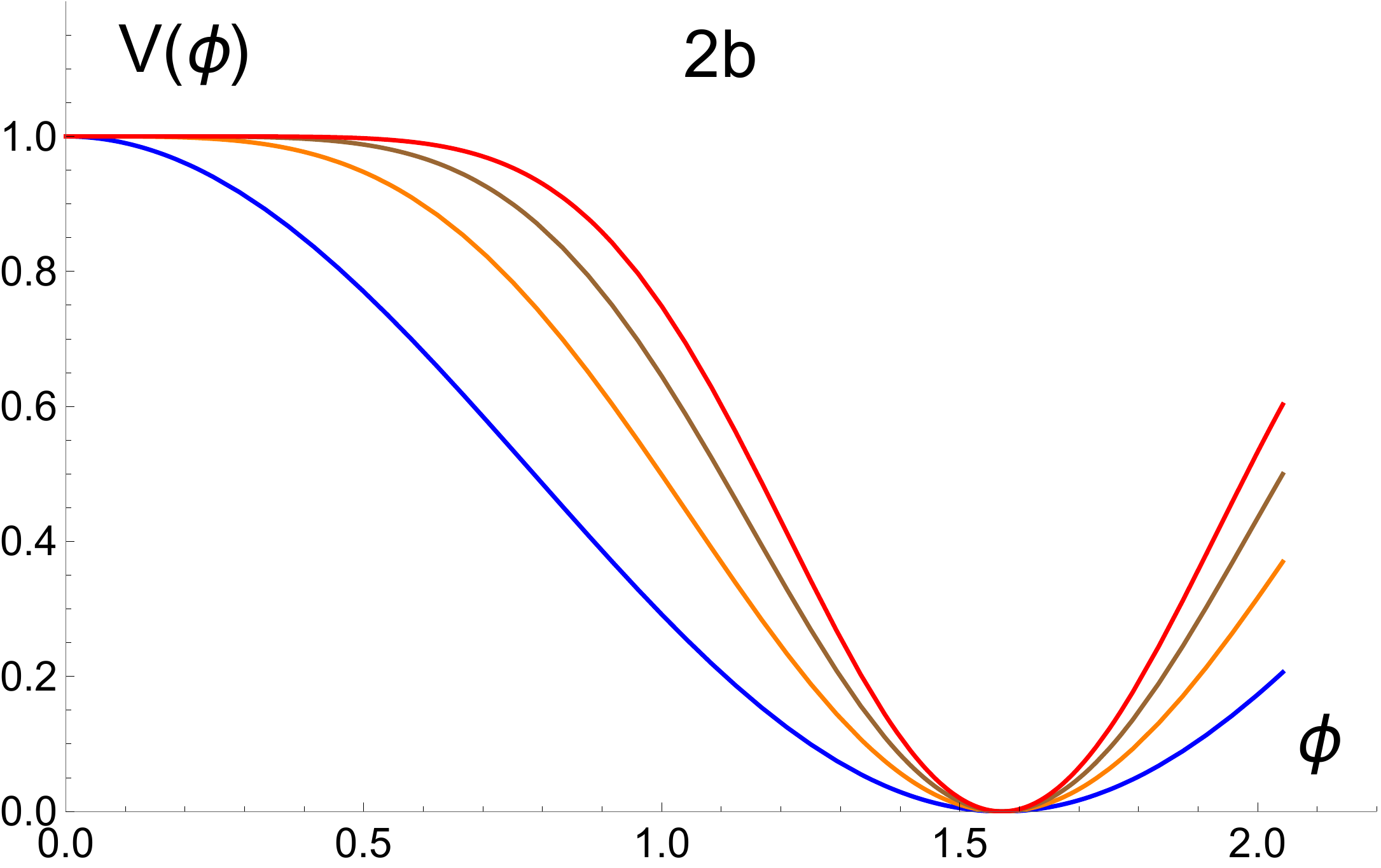}
\end{center}
\caption{In the figures above $V_0$ and $f$ have been arbitrarily set equal to 1 for illustrative purposes only. Panels $1a$ and $1b$ show the potentials \eqref{potodd} and \eqref{poteven} for $p=3,5,7,9$ and $p=2,4,6,8$, respectively. The original versions of the Natural Inflation potential \eqref{potNI} are also shown by the dashed lines. Panels $2a$ and $2b$ show the same potentials when they have been displaced to the left by $f\pi/2$ and the range for $\phi$ is now restricted to $0 < \phi < f \pi/2 $. As a result both potentials can now be represented by a single formula as given by  Eq.~\eqref{pot} and shown by  Fig.\,\ref{potfig} below.}
\label{NIi}
\end{figure}
An undesirable feature of single field NI is that the scale $f$ typically takes values greater than one (in Planck units) making an interpretation of the model in terms of effective field theory difficult. The value of $f$ is set by the requirement of having a sufficient number of e-folds of inflation.
As we will see, in the following model $f$ can take practically any value with values less than one favored by the requirements of having inflation consistent with the observations. We could try modifying the NI potential to the form
\begin{equation}
V= V_0\left(1- \cos\left(\frac{\phi}{f}\right)\right)^p, 
\label{potMu}
\end{equation}
however, it is easy to see that for most values of $p$ it would be difficult to have a simple interpretation in terms of inflation and, furthermore, $f$ would still be greater than one.

\begin{figure}[tb]
\begin{center}
\includegraphics[trim = 0mm  0mm 1mm 1mm, clip, width=6.5cm, height=5cm]{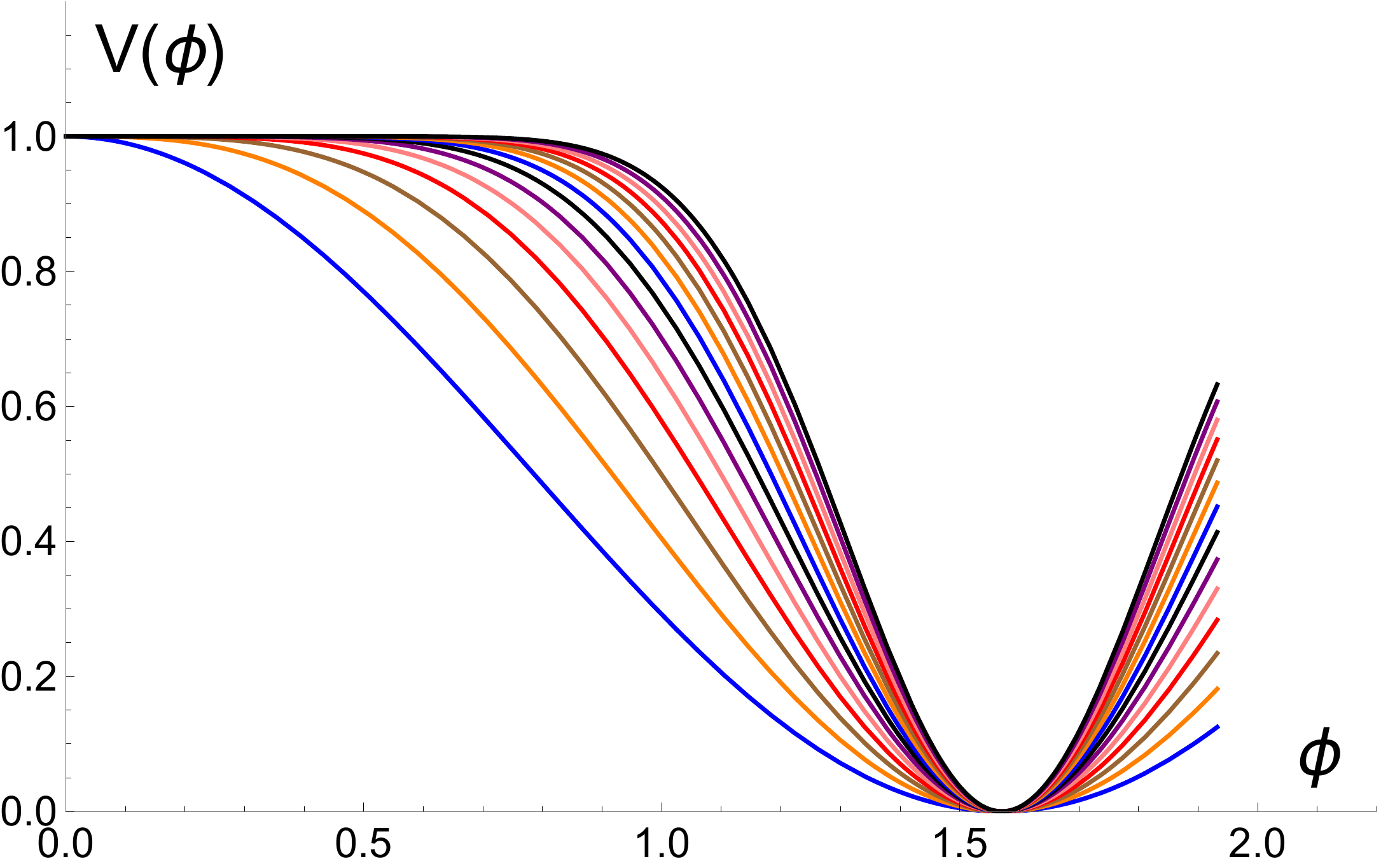}
\includegraphics[trim = 0mm  0mm 1mm 1mm, clip, width=6.5cm, height=5.cm]{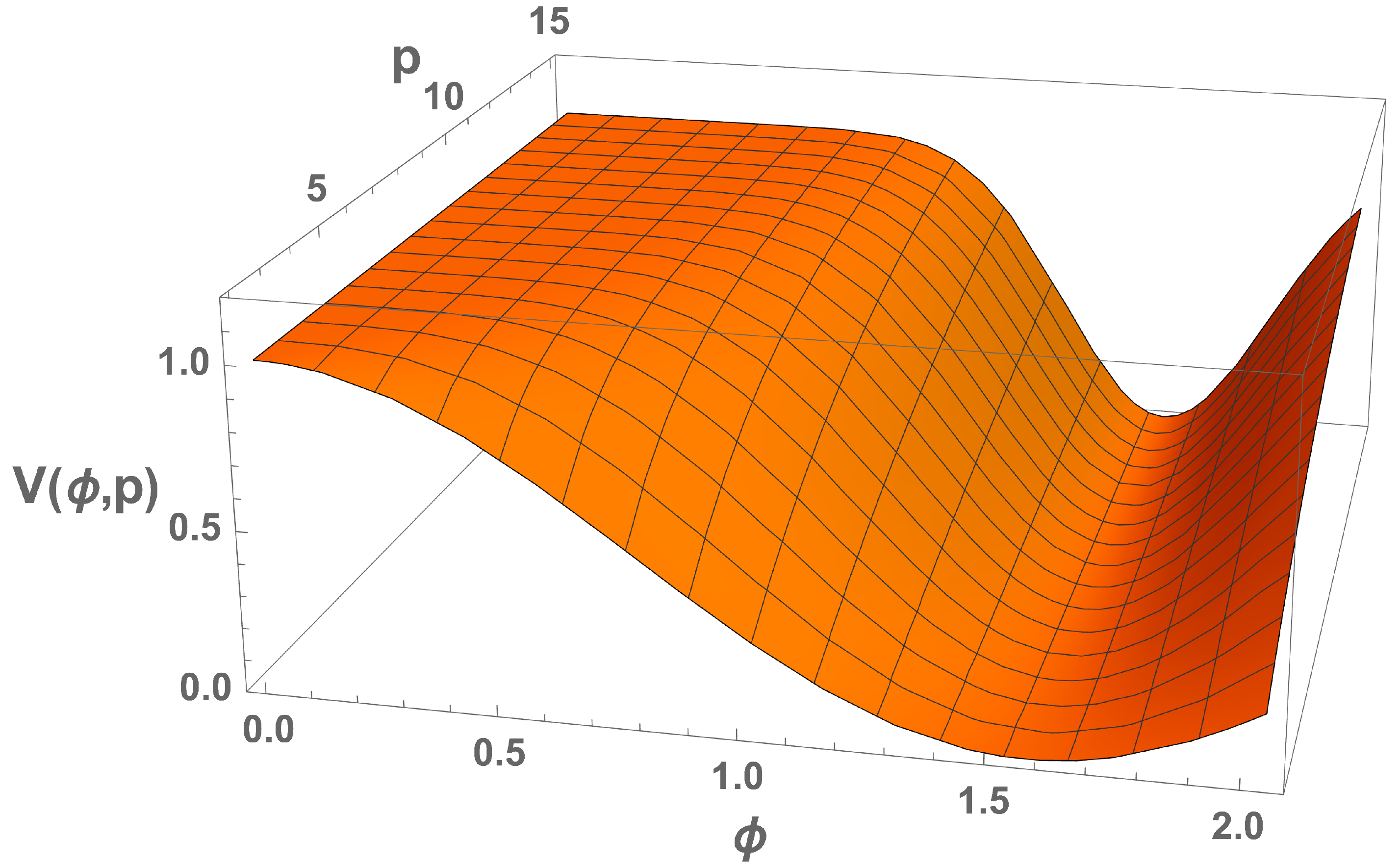}
\caption{\small  The l.h.s figure shows the potential \eqref{pot} for $p=2, 3, 4, ..., 15$ (from bottom to top) while the r.h.s figure shows the same potential with a continuously varying parameter $p$. The relevant range for observable inflation to occur is contained in the interval $0<\phi<f\pi/2$, as shown for the $f=1$ case. For $p>2$ the potential is flat at the origin with vanishing first and second derivatives.
}
\label{potfig}
\end{center}
\end{figure}

\begin{figure}[tb]
\begin{center}
\includegraphics[width=9cm]{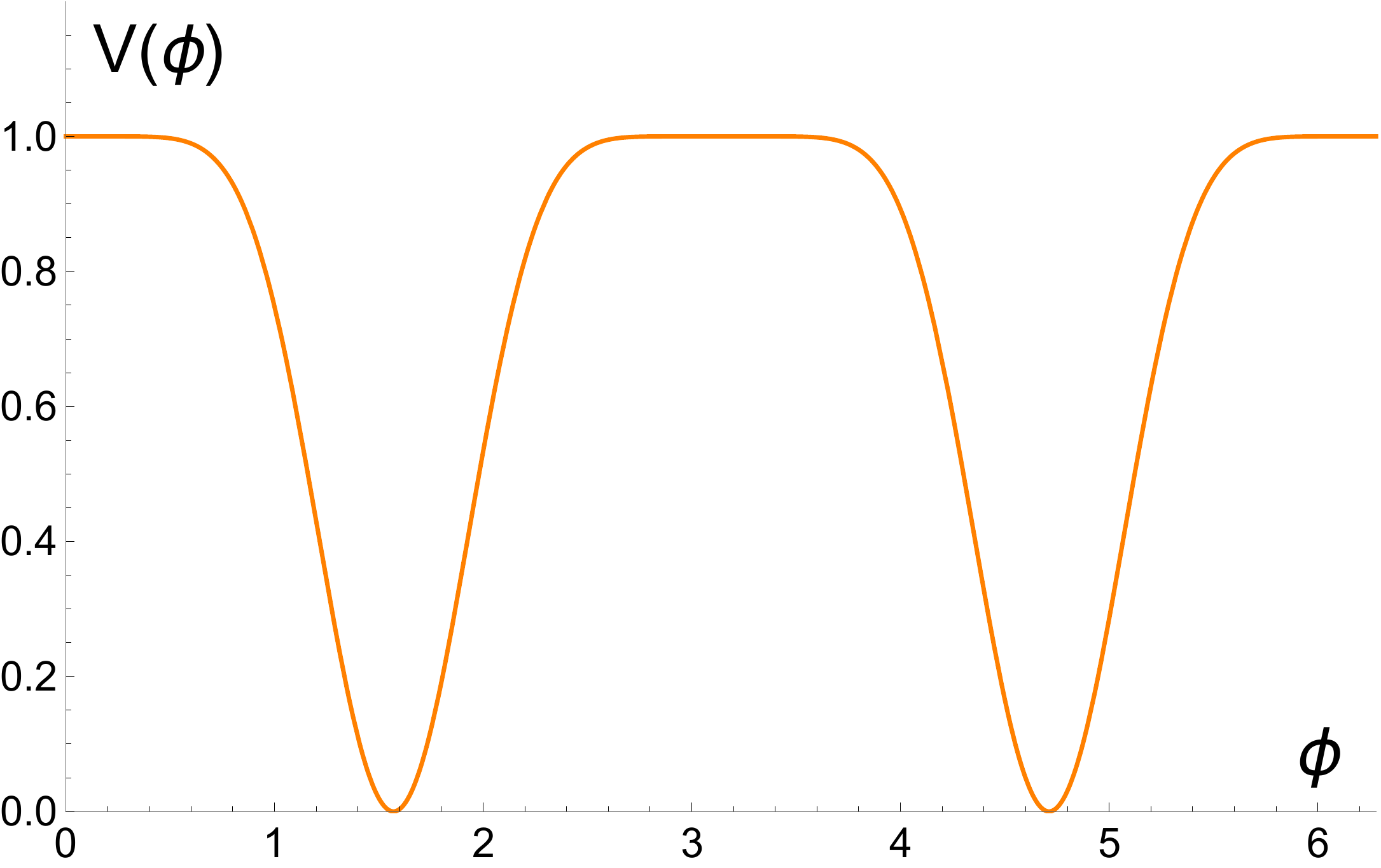}
\caption{\small Perhaps the most attractive expression of the potential \eqref{pot} is with even-$p$ as illustrated in the figure above for an extended range which shows its periodicity in two cicles and for the $p=8$ case. The minimum has always a leading term of the $\phi^2$-type while the plateau is proportional to $\phi^p$, see discussion of this point after Eqs.~\eqref{potexporigin} and \eqref{potexpmin}.
}
\label{veven}
\end{center}
\end{figure}

Usually the Natural Inflation (NI) potential is expressed in one of the following two equivalent forms
\begin{equation}
V_{NI} = V_0\left(1\pm \cos\left(\frac{\phi}{f}\right)\right).
\label{potNI}
\end{equation}
We modify each of the two versions of the NI potential as follows
\begin{equation}
V= V_0\left(1+ \cos^p\left(\frac{\phi}{f}\right)\right),  \quad\quad\quad (p-odd)
\label{potodd}
\end{equation}
and
\begin{equation}
V = V_0\left(1- \cos^p\left(\frac{\phi}{f}\right)\right).  \quad\quad\quad (p-even)
\label{poteven}
\end{equation}
These potentials are shown in Fig.\,\ref{NIi}, panels $1a$ and $1b$ and compared with the original potential of Eq.~\eqref{potNI} (dashed line). Doing a translation by $f\pi/2$ the potential for the $p$-odd case goes to (see Fig.\,\ref{NIi}, panel $2a$)
\begin{equation}
V = V_0\left(1+ \cos^p\left(\frac{\phi}{f}\right)\right)  \rightarrow  V_0\left(1+ \cos^p\left(\frac{\phi+f\pi/2}{f}\right)\right)=V_0\left(1+\left(-\sin\left(\frac{\phi}{f}\right)\right)^p\right),
\label{potoddred}
\end{equation}
while for $p$-even (see Fig.\,\ref{NIi}, panel $2b$)
\begin{equation}
V = V_0\left(1- \cos^p\left(\frac{\phi}{f}\right)\right)  \rightarrow  V_0\left(1- \cos^p\left(\frac{\phi+f\pi/2}{f}\right)\right)=V_0\left(1-\left(-\sin\left(\frac{\phi}{f}\right)\right)^p\right).
\label{potevenred}
\end{equation}
In both cases the resulting potential is 
\begin{equation}
V = V_0\left(1-\sin^p\left(\frac{\phi}{f}\right)\right),\quad\quad\quad  (p>1,\quad 0 < \phi < f \pi/2 )
\label{pot}
\end{equation}
and it is shown in the l.h.s of Fig.\,\ref{potfig} for $p=2, 3, 4, ..., 15$ while the r.h.s figure shows the same potential with a continuously varying parameter $p$. The relevant range for observable inflation to occur is contained in the interval $0<\phi<f\pi/2$, as shown. Thus, in what follows we study the potential given by Eq.~\eqref{pot} exclusively. For $p>2$ this potential is flat with vanishing first and second derivatives at the origin. Perhaps the most attractive expression of the potential \eqref{pot} is with even-$p$ as illustrated in Fig.\,\ref{veven} for an extended range which shows its periodicity and for the $p=8$ case.

An expansion of the potential \eqref{pot} around the origen is 
\begin{equation}
V/V_0 = 1-\left(\frac{\phi}{f}\right)^p+\frac{p}{6}\left(\frac{\phi}{f}\right)^{p+2}+\cdot\cdot\cdot,
\label{potexporigin}
\end{equation}
while at the minimum
\begin{equation}
V/V_0 =\frac{p}{2}\left(\frac{\phi-\pi f/2}{f}\right)^2-\frac{p(3p-2)}{24}\left(\frac{\phi-\pi f/2}{f}\right)^4+\cdot\cdot\cdot.
\label{potexpmin}
\end{equation}
Thus, we see a strong  dependence on $p$ at the origin while at the minimum $p$ is only a proportionality constant to the leading $\phi^2$-term; the higher the power $p$, the flatter the potential (see Fig.\,\ref{veven}).  
Although the equation \eqref{pot} looks simple, inflation with this potential is very difficult to study analytically. So for now we will content ourselves by  giving some numerical examples by means of figures that show the viability of the model, pending an extensive numerical study. We could try to study the model analytically by working with an expansion of the potential as given by equations \eqref{potexporigin} and \eqref{potexpmin}. We could attempt to find the value of the inflaton at horizon crossing $\phi_k$ by solving \eqref{potexporigin}  and the end of inflation with the help of \eqref{potexpmin}. However, even in this simplified scenario, it is not possible to solve for arbitrary $p$ having to study particular values of $p$. But because the data match is obtained for a relatively large $p$ ($p \geq 6$), the resulting solutions are intractable, which unfortunately rules out an analytical approach based on the approximate expressions \eqref{potexporigin} and \eqref{potexpmin}.
It is also clear from the figures that the parameter $f$, which in the original NI model is associated with the scale of spontaneous symmetry breaking, can naturally take values less than one (in Planck units). Also the field variation $\Delta\phi$ is, in all cases studied, less than one as long as $f\lessapprox 1$, which is satisfied without any difficulty.
\section {\bf Viability of the model}\label{VIA} 

Below we present numerical results through various figures that show the feasibility of the model to produce inflation consistent with the most recent observations. In the absence of an analytical treatment of the various inflationary characteristics, for now we will content ourselves with showing that the model is capable of satisfying  bounds for the observables and at the same time producing an acceptable number of e-folds. 

\begin{figure}[tb]
\begin{center}
\includegraphics[trim = 0mm  0mm 1mm 1mm, clip, width=10cm, height=7cm]{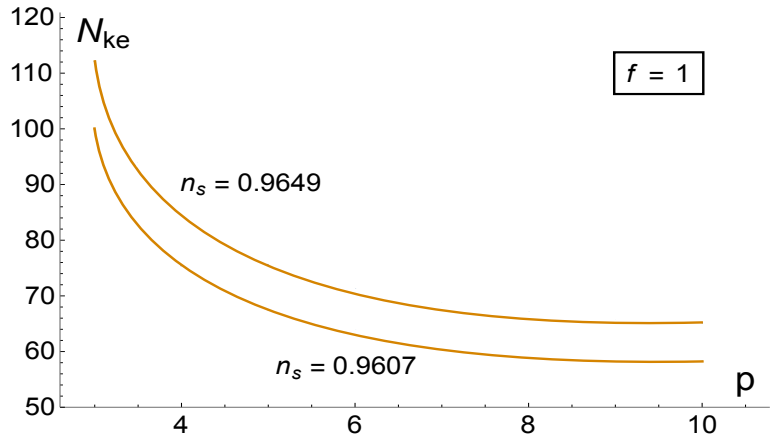}
\caption{\small The number of e-folds during inflation $N_{ke}$  as a function of $p$ for the lower bound of the spectral index $n_s=0.9607$ and also for the central value $n_s=0.9649$, as reported by Planck  \cite{Akrami:2018odb} . The plot corresponds to $f=1$, lines for lower values of $f$ are not reproduced because they are very similar to the ones shown above. For $f=10^{-3},..., 10^{-10}$ maximum variations on $N_{ke}$ amount to a mere 2 e-folds downwards for large $p$. Thus, $N_{ke}$ is practically $f$-independent.
}
\label{Nke}
\end{center}
\end{figure}

In Fig.~\ref{Nke} we give the number of e-folds for the lower bound of Planck \cite{Akrami:2018odb}  reported values for the spectral index $n_s=0.9607$ and also for the central value $n_s=0.9649$. This is done for  $f=1$ for $p=3$ to 10 and calculated for $p=10^2,10^3$ reaching in these last two cases values of $N_{ke}$ close to 50, in the $n_s=0.9607$ case. We have also done the same figure for  $f=10^{-3}, 10^{-6}$ and $f=10^{-8}$ and find that no big changes occur for $N_{ke}$ differing at most by 2 e-folds down from the results shown here for the $f=1$ case. Thus  $N_{ke}$ is practically $f$-independent but it is clearly $p$-dependent, we also see that for $p\gtrapprox 8$, $N_{ke}\leq 60.$ As $n_s$ grows from its lower value so does $N_{ke}$.

\begin{figure}[tb]
\begin{center}
\includegraphics[trim = 0mm  0mm 1mm 1mm, clip, width=10cm, height=7.5cm]{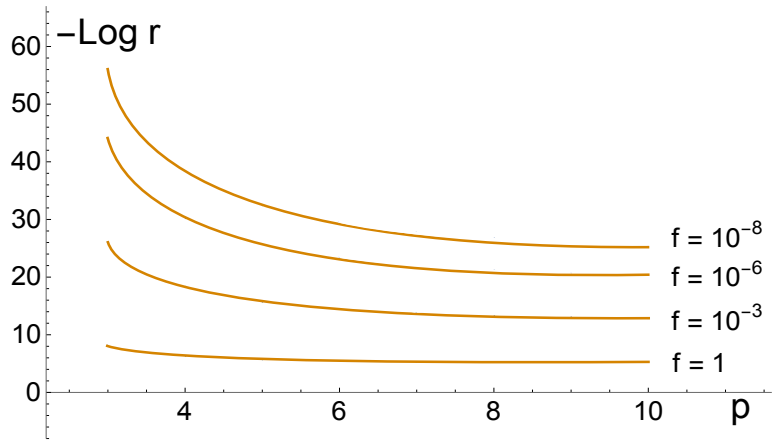}
\caption{\small Plot of (minus) the logarithm of the tensor-to-scalar ratio $r$ as a function of $p$ for various values of the symmetry breaking scale $f$ and for the lower bound $n_s=0.9607$. The highest value of $r$ occurs here for $f=1$ and is of the order of $10^{-6}$.
}
\label{rdep}
\end{center}
\end{figure}

In Fig.~\ref{rdep} we show the behavior of the tensor-to-scalar ratio as a function of $p$ for $f=1, 10^{-3}, 10^{-6}, 10^{-8}$ for the lower bound $n_s=0.9607$. The ratio $r$ typically takes small values with the largest (when $f=1$) of the order of $10^{-6}$ becoming smaller with $f$.

\begin{figure}[tb]
\begin{center}
\includegraphics[trim = 0mm  0mm 1mm 1mm, clip, width=10cm, height=7.5cm]{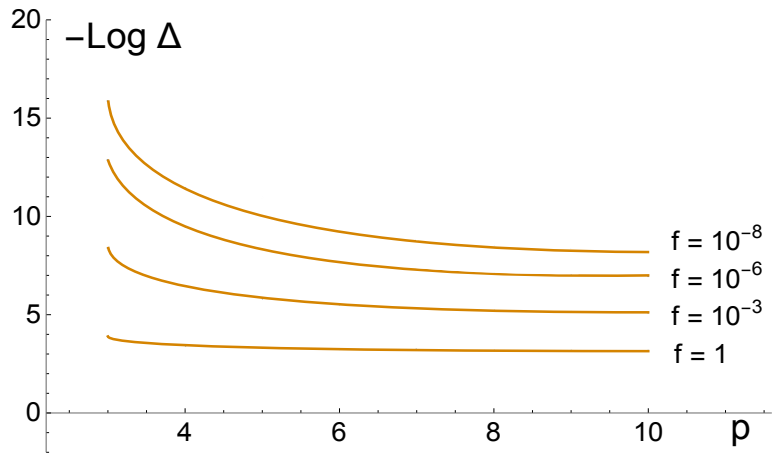}
\caption{\small  Plot of (minus) the logarithm of the inflationary scale $\Delta\equiv V_k^{1/4}$ as a function of $p$ for $f=1, 10^{-3}, 10^{-6}, 10^{-8}$ for the lower bound $n_s=0.9607$. The highest value of $\Delta$ occurs here for $f=1$ and is of the order of $10^{-4}$ in Planck units.
}
\label{Deltadep}
\end{center}
\end{figure}

In Fig.~\ref{Deltadep} we show the behavior of the scale of inflation $\Delta\equiv V_k^{1/4}$ as a function of $p$ for $f=1, 10^{-3}, 10^{-6}, 10^{-8}$ for the lower bound $n_s=0.9607$. The scale $\Delta$ takes typically small values with the largest (when $f=1$) of the order of $10^{-4}$ (in Planck units) becoming smaller with $f$.

\begin{figure}[tb]
\begin{center}
\includegraphics[trim = 0mm  0mm 1mm 1mm, clip, width=7.5cm, height=6.cm]{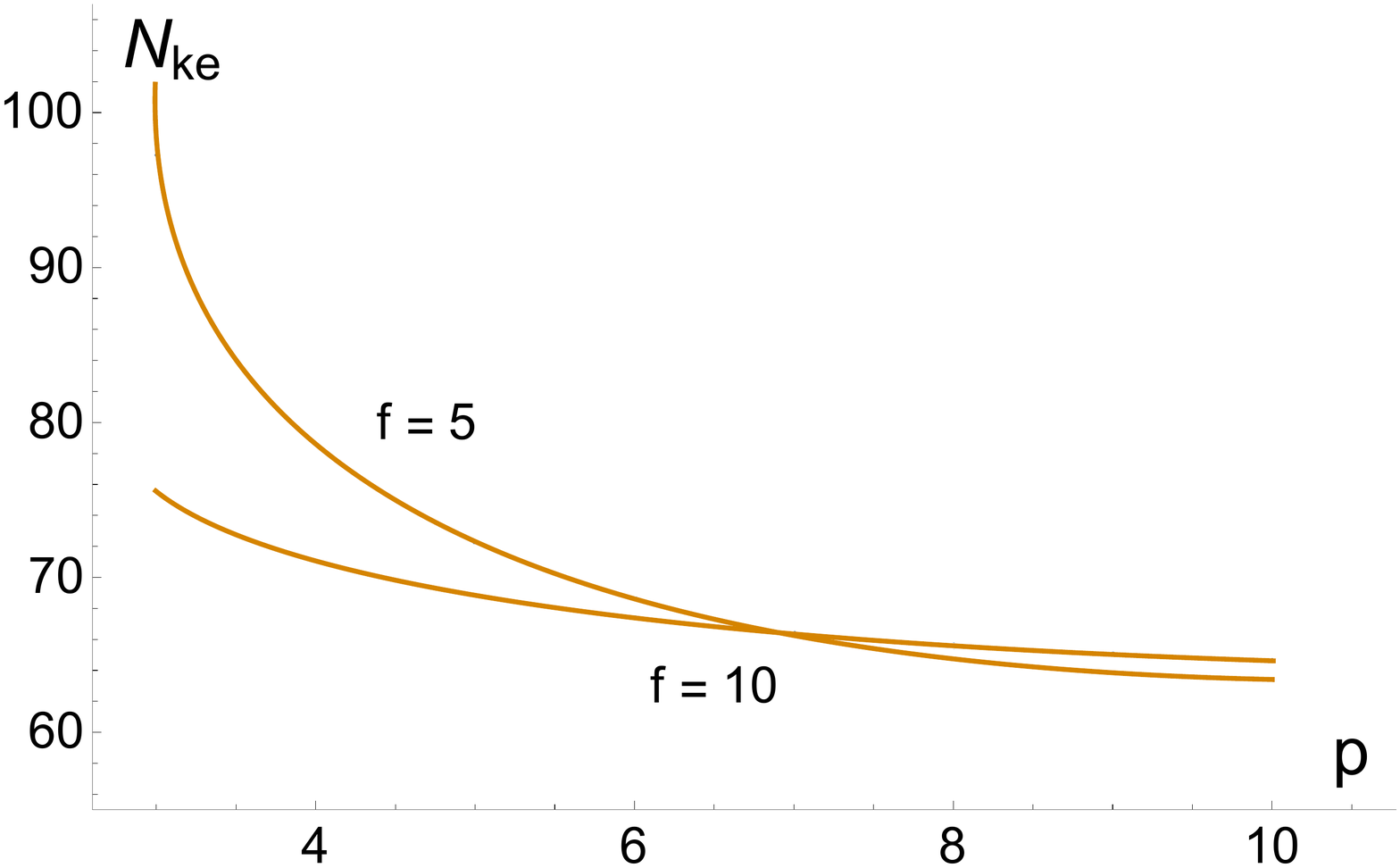}
\includegraphics[trim = 0mm  0mm 1mm 1mm, clip, width=7.5cm, height=6.cm]{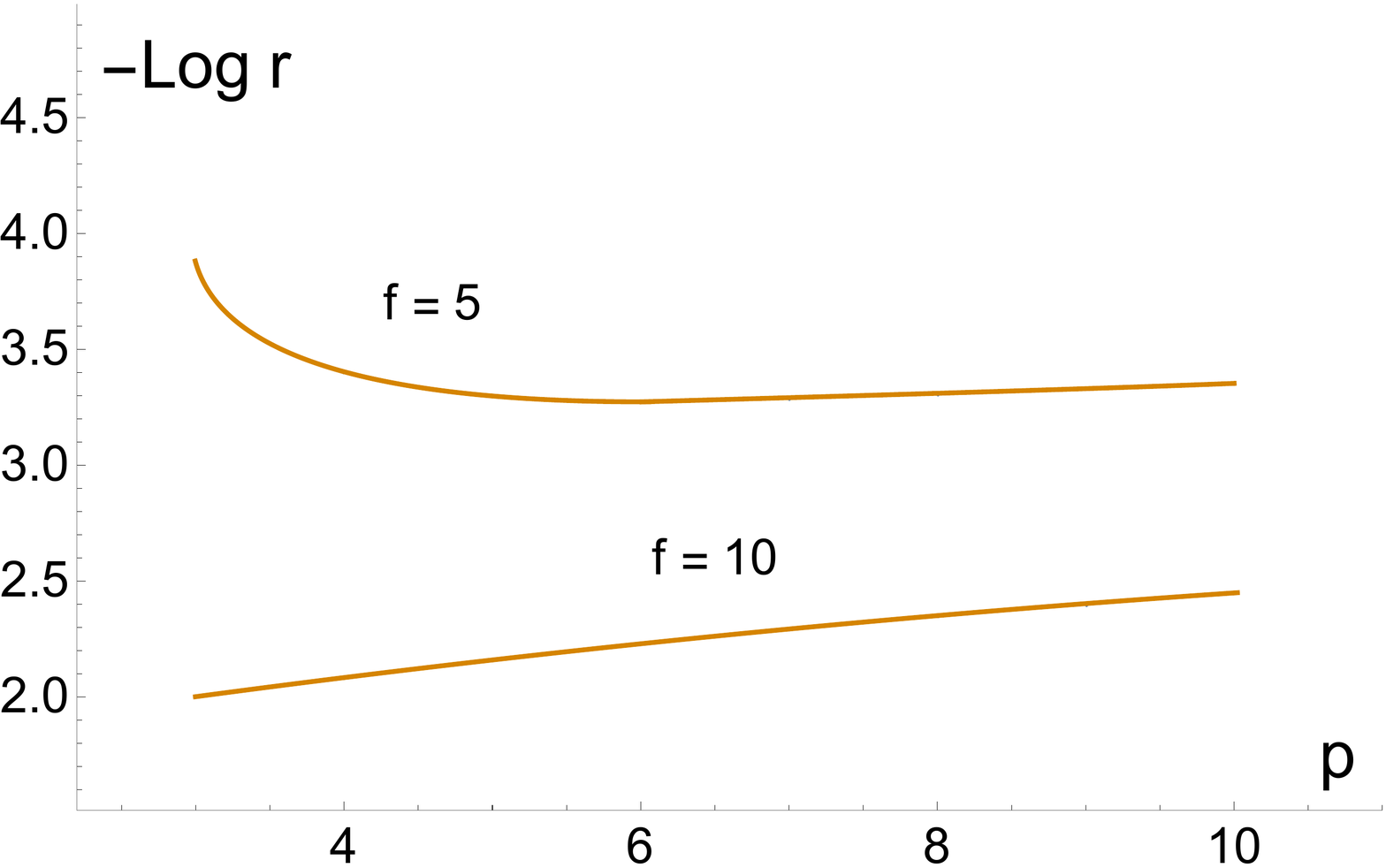}
\includegraphics[trim = 0mm  0mm 1mm 1mm, clip, width=7.5cm, height=6.cm]{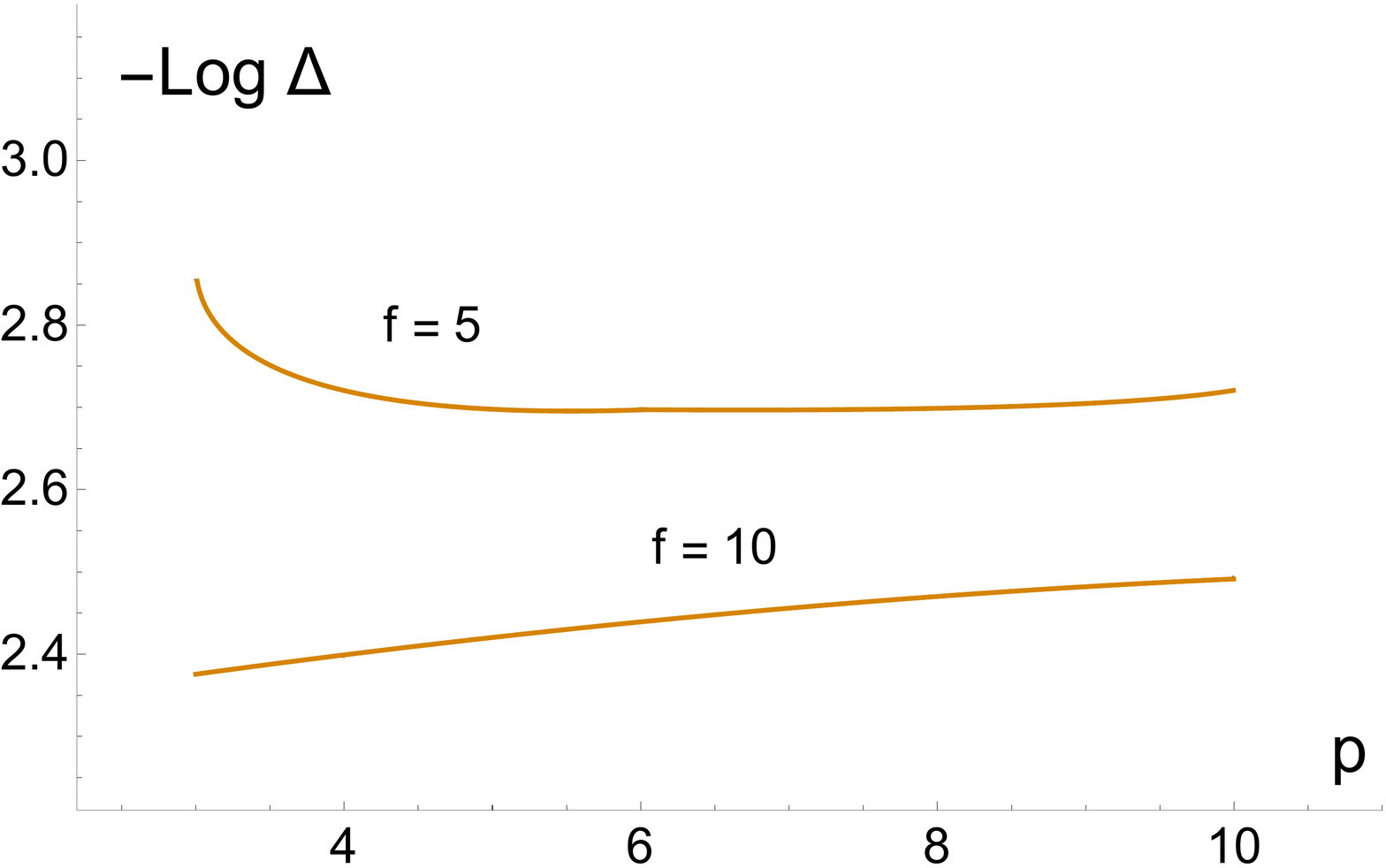}
\end{center}
\caption{Plot of the number of e-folds $N_{ke}$, (minus) the logarithm of the ratio $r$ and the scale of inflation $\Delta\equiv V_k^{1/4}$ as functions of $p$ for two values of $f$ larger than one. We see that desirable values for the number of e-folds $N_{ke} \leq 60$ are not reached suggesting that $f>1$ is not favored.}
\label{Nke510}
\end{figure}

Finally we show in Fig.~\ref{Nke510} the same quantities as above for two values of $f$ larger than one: $f=5,10$  for the lower bound $n_s=0.9607$. We see that $N_{ke}$  becomes larger than 60 e-folds even for the lower value of $n_s$ suggesting that $f>1$ is less favored than the cases with $f\leq 1$.

\section {\bf Conclusions}\label{CON}

We have studied the Natural Inflation (NI) model with emphasis on its transition to a power law potential. We have shown explicitly how the NI potential transitions smoothly, in the $f\rightarrow\infty$ limit, to the monomial $\phi^2$. This explains why, in the $n_s$-$r$ plane, the lines $r(n_s, N_{ke})$ for varios $N_{ke}$ end in the monomial $\phi^2$. On the other hand the fact that the symmetry breaking scale $f$ turns out to be greater than one (in Planck units) has been an unsatisfactory issue of the model. Thus, a modification of NI is proposed such that in the resultant model $f$ can take values smaller than one and at the same time allows a satisfactory number of e-folds. The proposed model satisfies all the inflationary requirements imposed by the most recent observations.
\acknowledgments
We acknowledge financial support from UNAM-PAPIIT,  IN104119, {\it Estudios en gravitaci\'on y cosmolog\'ia}. 
\\
\\
All data generated or analysed during this study are included in this published
article.



\begin{thebibliography}{10}
\bibitem{Freese:1990rb}
K. Freese, J. A. Frieman,  and A.V. Olinto,
\newblock {Natural inflation with pseudo - Nambu-Goldstone bosons},
\newblock {Phys. Rev. Lett.} {\bf 65}, 3233-3236 (1990).

\bibitem{Akrami:2018odb} 
Y.~Akrami {\it et al.} [Planck Collaboration],
\newblock {Planck 2018 results. X. Constraints on inflation},
\newblock {Astron. Astrophys.} {\bf 641}, A10 (2020).

\bibitem{Knox:1992iy}
L.~Knox, M.S.~Turner,
\newblock {Inflation at the electroweak scale},
\newblock {Phys. Rev. Lett.} {\bf 70}, 371-374 (1993).

\bibitem{Kinney:1995cc}
W.H.~Kinney,  and K.T.~Mahanthappa,
\newblock {Inflation at low scales: General analysis and a detailed model}.
\newblock {Phys. Rev. D}  {\bf 53}, 5455-5467 (1996).

\bibitem{Kinney:1995ki}
W.H.~Kinney,  and K.T.~Mahanthappa,
\newblock {Inflation from symmetry breaking below the Planck scale},
\newblock {Phys. Lett. B} {\bf 383}, 24-27 (1996).

\bibitem{German:2021abc} 
G. Germ\'an,
\newblock {New generalization of the simplest $\alpha$-attractor $T$ model},
\newblock {Phys. Rev. D} {\bf xx}, xxxxxxxx (2021). To appear.

\bibitem{German:2021tqs} 
G.~Germ\'an,
\newblock {On the $\alpha$-attractor $T$ models},
\newblock {J. Cosmol. Astropart. Phys.} 09 (2021) 017.

\bibitem{Kallosh:2013yoa}
R. Kallosh, A. Linde and D. Roest,
\newblock {Superconformal Inflationary $\alpha$-Attractors},
\newblock {J. High Energy Phys.}  {\bf 11}, 198 (2013).
 
\end{thebibliography}
\end{document}